\documentstyle{l-aa}
\input epsf.sty
\newcommand{\be}{\begin{eqnarray}}
\newcommand{\ee}{\end{eqnarray}}
\newcommand{\bew}{\begin{eqnarray*}}
\newcommand{\eew}{\end{eqnarray*}}
\newcommand{\eq}{\begin{equation}}
\newcommand{\en}{\end{equation}}
\newcommand{\msun}{M$_\odot$}

\def\teff{T$_{\rm eff}$}

\def\teff{T$_{\rm eff}$}

\def\plottwo#1#2{\centering \leavevmode
\epsfxsize=.45\textwidth \epsfbox{#1} \hfil
\epsfxsize=.45\textwidth \epsfbox{#2}}
\begin{document}
\thesaurus{ 08.01.1 - 08.01.3 - 08.03.1 - 08.16.3 - ?? }
\title{Spectral evolution of and radiation energy generation
by coeval stellar
populations with different initial composition and chemical enrichment}

\author{Peeter Traat}
\institute{Tartu Astrophysical Observatory, 61602 T\~oravere, Estonia}
%
%
\date{Received date; accepted date}
\maketitle
\markboth{Peeter Traat: Evolution of stellar populations with 
different $Z$ and chem. enrichment}{}

\begin{abstract}
Initial chemical composition of stars is, besides their mass, another key  
factor in stellar evolution. Through stellar lifetimes and impact on 
radiation output and nucleosynthesis of stars it is controlling both the pace 
of evolution of galactic matter/light and changes in their integrated 
observables and spectra. However, because of long-time scarcity of homogeneous
stellar evolutionary data for compositions other than solar, or "normal",
all the estimates of physical situation and global parameters for galaxy
collective (radiation density, "global" star formation rate (SFR) etc.) have
effectively remained being restricted to the assumption of normal composition
and the composition-dependence of galactic properties ignored even when it
should be obviously rather significant, e.g. in discussions of the main
galaxy-formation epoch between $z \sim 1\div 10$.

With our evolutionary synthesis codes we have performed extensive 
computations of temporal behaviour of photometric and spectral properties of
stellar populations having different initial $Z$ and varying IMF-s and SFR-s.
Also, the production of most common nucleosynthesis elements He, C and O was
followed. These computations have been performed on the basis of two available
but different homogeneous  multicomposition stellar evolution tracks grids by
Geneva and Padova groups and the Kurucz model atmospheres (Kurucz 1993).

In given paper next to the discussion of overall effects of evolutionary
differences we also present and comment the normalized per stellar mass
unit standard tables of the detailed  radiation energy output from stellar
populations with different chemical  composition, integrated over the whole
lifetime of their stars, likewise the tables on He, C and O production. The
energy production at different  wavelengths per production of unit amount of
$^{16}$O and heavy elements are tabulated in the function of the "metallicity"
$Z$, the slope parameter $n=x+1$ (sic!) of the initial mass function (IMF) and
the lowest stellar initial mass boundary, $M_{\rm cut}$, for stars evolving 
finally into black holes.

\end{abstract}

\section{Introduction}

Apart from the initial nucleosynthesis in the first $\sim 10^3$ seconds after 
the Big Bang, what settled the primordial abundances of light elements (cf. 
Wagoner et al. 1967; Walker et al. 1991; Smith et al. 1993), the only source 
of the chemical enrichment  of the matter in the 
post-recombination Universe has been the nucleosynthesis by stars, with its 
products mixed into the ambient medium by stellar ejecta 
during rapid mass loss stages and through stellar winds.
This way stellar nucleosynthesis has been the 
major driving force behind all the subsequent evolution of stellar systems 
 and the main generator of radiation energy and the cosmic background  
 from the formation of 
very first stars up to the present epoch. 

As known, rapid nucleosynthesis in hot matter following the Big Bang settles
the initial He content, $Y_0$, and produces negligible amounts of $^{3}$D, 
$^{6}$Li, $^{7}$Li, Be $etc.$  Cosmology cannot tell us about the exact time
and redshift, when the first stars might have appeared in the Universe and
therefore the first stellar systems started to form, but it is believed to have
happened at
redshifts  around $z\,\sim \,10$ or even somewhat higher. Originally extremely
deficient in heavy elements, the chemical  composition of the matter of these
primeval galaxies will evolve rather  unsignificantly in low-mass low-density
objects, but in the high-mass objects  the enrichment proceeds rapidly.
Consequently, the chemical composition  differences should be of great  
importance, when the radiation of young galaxies is being analyzed or used to
address the cosmological problems.  

\section{The era of first galaxies}

Longstanding observational search for the primeval galaxies, going
historically back to a paper by Partridge and Peebles (1967), has by now
pushed the redshift limit for the directly observed starforming objects the
farther and farther, with the most distant objects presently known having
$z\sim 7$. Although seemingly quite rare objects at so high redshifts, the
discovery of quasars at $z > 6$ (presumably surrounded by stellar component
and fainter galaxies) testifies, that galaxy formation process and birth of
first stellar populations is in some cases going very intensively on even in
that young the Universe, with age slightly less than 1 Gy. For the most 
distant quasar presently known, J1148+5251 with redshift of $z=6.419 $ (Fan et
al. 2003), Bertoldi et al (2003) give on the basis of measured dust emission
for its parent protogalaxy an estimate of very impressive star formation, with
the rate $\sim$3000 \msun/yr. Many far-away galaxies have been discovered in
quasar fields, some of them seem even to be rather evolved. For example, deep
imaging  with Keck telescope of the field around PC 1247+3406  (Soifer et al.
1994), having $z = 4.897$, and the $z = 3.8$ radio galaxy 4C 41.17  (Graham et
al.  1994) have revealed the hints of a population of red galaxies at
magnitudes  $K \sim 21$ and fainter. Graham et al. point out that the closest
companions  to the 4C 41.17 are very red with their emission most likely
caused by  starlight, with lower limit to their age deduced from initial
starburst 0.5  Gyr, and if sharing the same redshift with 4C 41.17, should
have formed  at $z\,${\tiny $\stackrel{>}{\sim}\,$}8.

Another but indirect evidence of the ongoing star/galaxy formation at these 
and even higher redshifts is the absence of continuum absorption shortward 
of Ly$\,\alpha$ in quasar spectra (Gunn-Peterson effect), what testifies that 
the intergalactic medium 
has been 
practically completely ionized starting from the very early 
epochs down to the present. This is generally attributed to the 
stellar/galactic  
processes be they supernova-driven winds either from luminous (Schwarz et al. 
1975) or low-mass (Tegmark et al. 1993) high-$z$ young 
galaxies, cosmic rays (Ginzburg and Ozernoy 1965; Nath and Biermann 1993) or 
the photoionization by stellar (Songaila et al. 
1990; Miralda-Escud\'e and Ostriker 1990) or quasar 
(Arons and McCray 1969) UV-radiation. 
Of course, in reality all these sources get combined in producing the
observed outcome. More exotic explanations have also included 
cosmological blast waves (Ostriker and Ikeuchi 1983), decaying massive 
neutrinos (Sciama 1990) etc. 

First discovered by Gunn and Peterson (1965),
the constraints to the effect got further developed by a number of 
observers, e.g. by Giallongo et al. (1994). They deduced, measuring 
the average depression between Lyman absorption lines of the quasar 
BR 1202-0725 ($z_{em} = 4.695$) the value of optical depth 
$\tau \leq 0.02 \pm 0.03$ at $z \simeq 4.3$ and, taking the quasars to be 
the prime ionizators of the gaseous medium and maximizing their known 
statistics, the estimate for the density of the IGM $\Omega_{IGM}\,${\tiny 
$\stackrel{<}{\sim}\,$}0.01. Juxtaposition of that result with the mean 
baryonic density deduced either from HI Ly$\,\alpha$ absorption at
$z=1.9$, $\Omega_b = 0.044$ (Tytler et al. 2004), or from D/H measurements in
nearly pristine gas clouds on the basis of conventional Big Bang
nucleosynthesis, $\Omega_b = 0.0214\pm 0.0020\, h_{100}^{-2}$  (Kirkman et al.
2003, for the popularly accepted value $H_0=71$ $km
s^{-1}$Mpc$^{-1}$ it gives $\Omega_b = 0.0425\pm 0.004$) indicates, that even
at these very early times the  dominant part of the baryonic matter had been 
confined into dense  substructures, what occupied a rather limited fraction of
the total volume.  {\em Vice versa,} identifying these dense structures at
least partly being  young/primeval galaxies contributing to the ionizing
UV-radiation at these and higher  redshifts, it would be possible to satisfy
the balance $\Omega_{IGM} + \Omega_{gal} + \Omega_{H I} = \Omega_b$, with
$\Omega_{gal}$ and $\Omega_{H I}$ denoting the densities of galactic  matter
and neutral hydrogen clouds, respectively. 

\section{Spectrophotometric models}

Integrated spectra are the main sources of information on the 
stellar content and interstellar medium 
 of external galaxies farther than the Local Group. Being primarily built up by
thermal  radiation of stellar component, the integrated spectrum depends on its
age, star formation rate,  stellar mass spectrum and chemical composition
distribution, but  may also be heavily influenced by the reradiation and
absorption by dust and heated gas in starbursts. Owing to specific frequency
behaviour, the nonthermal component can be  easily recognized , 
 all the other factors work together in build-up of the spectrum in a complex 
 manner. 

 Evolutionary modelling of galactic {\it spectra} was started by Bruzual
(1983) more than two decades ago, but remained for most of that period
 mainly constrained to solar composition, due to both
lack   of homogeneous stellar evolutionary track sets for different
metallicities,as also underlined in the abstract, as well as spectral
libraries for nonsolar compositions. They became gradually available to the
midst of 1990's. Presently there do exist two consistent track sets with
sufficient range of metallicities -  set for 6 compositions produced by
Geneva group (cf. Meynet {\it et al.}  1994, and Charbonnel {\it et al.} 1996, 
with earlier references in these; a sixth, high-metallicity  $Z=0.10$ 
composition got added to the set later by Mowlavi {\it et al.} 1998)
and Padova tracks (
cf. Girardi {\it et al.} 1996, and references therein, Girardi {\it et al.}
2000 for stars of lower masses), the
latter covering with 8 compositions practically all the observable and
theoretically useful metallicity interval from  $Z=0.0001$ until $Z=0.10$.
Both yet still pose some smaller problems for  applications(ors) - 
Geneva tracks do not include for 4 non-solar compositions (except
$Z=0.001$) past-RGB stages of  evolution of low-mass stars $M \le 1.7$ \msun,
so these have to be added  from other 
(and inhomogeneous) sources, 
it also has shorter range towards extremely metal-deficient abundances.
Plus, the location of stars with masses $M < 0.8$ \msun on and subsequent
evolution from the zero-age main sequence is not available, it has to be
taken elsewhere. Padova tracks  are relatively free of so basic
problematics, except of  lack of stars $M>9$ \msun \, in $Z=0.10$ subset, 
some minor  irregularities arise also for $Z=0.001$  composition because they
were computed with different radiative opacities. 

The only set of model atmospheres, suitable for evolutionary spectral
syntesis, has been produced by Kurucz (1993), covering spectral range 90~\AA 
--   160 $\mu$m for twenty compositions. It is known that since Kurucz models
do not contain bands of some molecules, in the cool star region \teff $\le
4500$\raisebox{.6ex}{$\circ$}K improvements are needed, but being developed 
by a group associated with Uppsala, such more sound atmospheres for cool
stars have not became available yet.

At the end of 1970-s and beginning of 1980-s, I developed a code package for
computions of photometric evolution for stellar systems, using theoretical
tracks and observational color indices and bolometric corrections for stars of
different temperatures and luminosity classes. This package was somewhat
complemented at the beginning of 1990-s as to allow on the basis of stellar
atmospheres to compute spectral distributions for single populations as well
as for the case of continous star formation in model galaxies and to compute
the photometry for filters within their passbands. With this package, I have
performed a wide set of computations of  photometrical and spectral population
models with different chemistry and  input physics (incl. also variations of
stellar track sets). We have used both track sets in these  studies of spectral
evolution of stellar populations in function of their initial metallicity and
star formation  parameters. No gas emission/dust absorption has been
considered in the models, since it is highly individual for galaxies.  

Unfortunately, being made about ten years ago, the results of these
computions  have not been published in due fashion. On the basis of Padova
tracks and isochrones a standardized grid of spectral data files has been
released (Traat 1996) on a CD-ROM, produced as a collective effort to compile
a database for galaxy evolution modeling. The short description of that grid 
has been published in the paper, accompanying that CD-ROM and describing its
content (Leitherer {\it et al.} 1996).

That grid includes 1008 models and has the
widest practically useful   range of star formation
prescriptions. It    covers all the Padova set abundance range, i.e. eighth
metallicities   $Z = 0.0001$, 0.0004, 0.001, 0.004, 0.008, 0.02, 0.05 and
0.10. For
each $Z$ value that dataset includes  \begin{itemize} 
\item six power law IMF-s of different slopes $1.6 \div 3.5$, 
\item corresponding single-generation populations
(instantaneous formation, "initial starbursts")
\item     populations with continuous/continuing star formation for those
IMF-s      with 4 SFR index      values $s=0$ (constant SFR), $s=1$
(exponentially declining), $s=1.5$ and      2 (initially faster, later slower
than the exponential SFR) and 5      star-formation timescales 0.2, 1, 2, 5
and 15 Gy.   \end{itemize}
    Spectra were presented for 50 ages in the case of starburst populations 
and 20 ages for populations with continuous star formation. Ages range from 
0.003 Gy to 20 Gy. 

 SFR has been parametrized 
  by a power of the gas volume density (as introduced by Schmidt (1959)), with 
  index $s$ and time-scale $t_0$. In this context, the single-generation 
  ("initial-burst") populations with their independency on the SFR/its power 
  index form the limiting "starburst" $t_0 = 0$ case. 

\begin{figure*}

\plottwo{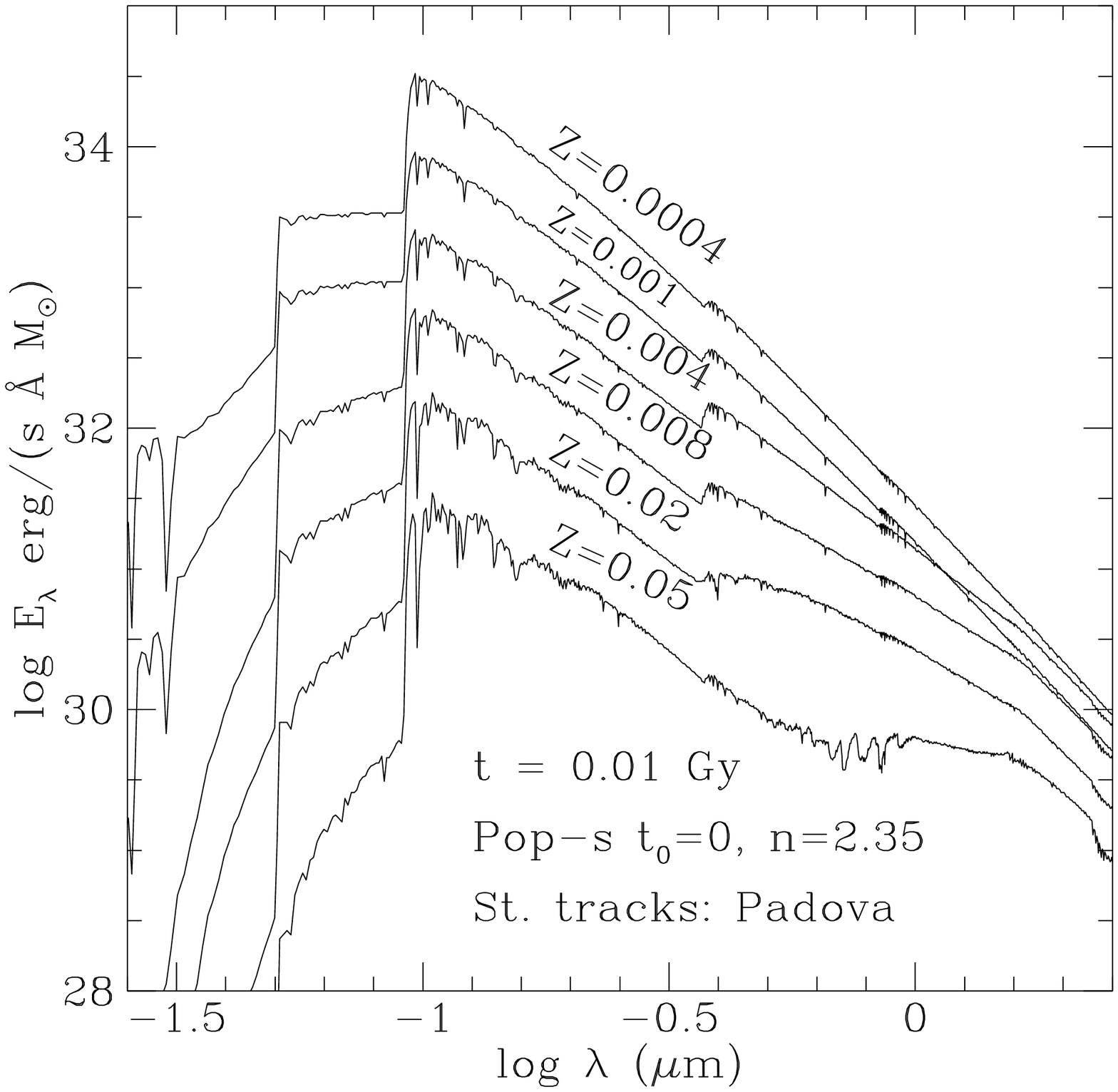}{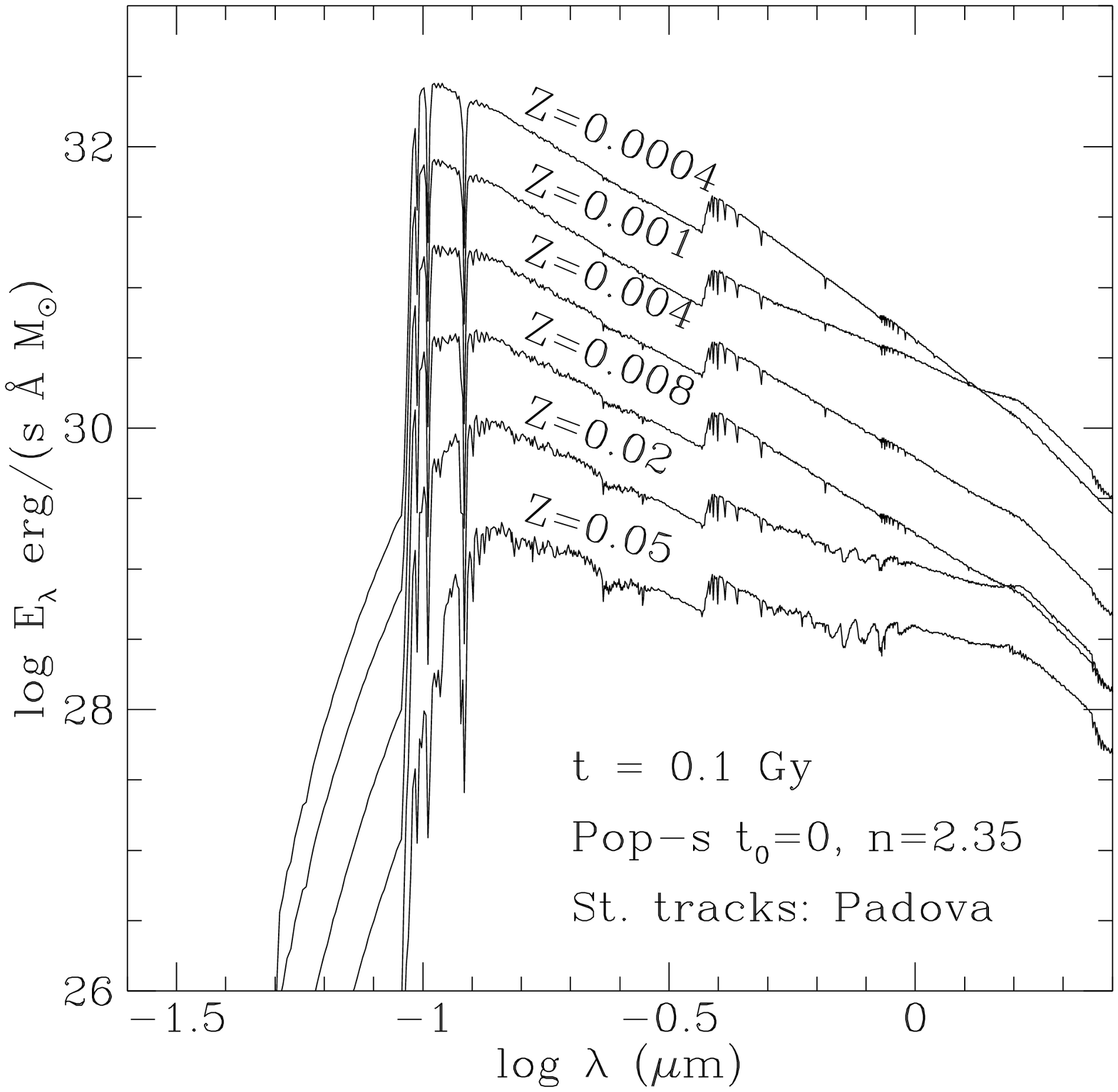}

\plottwo{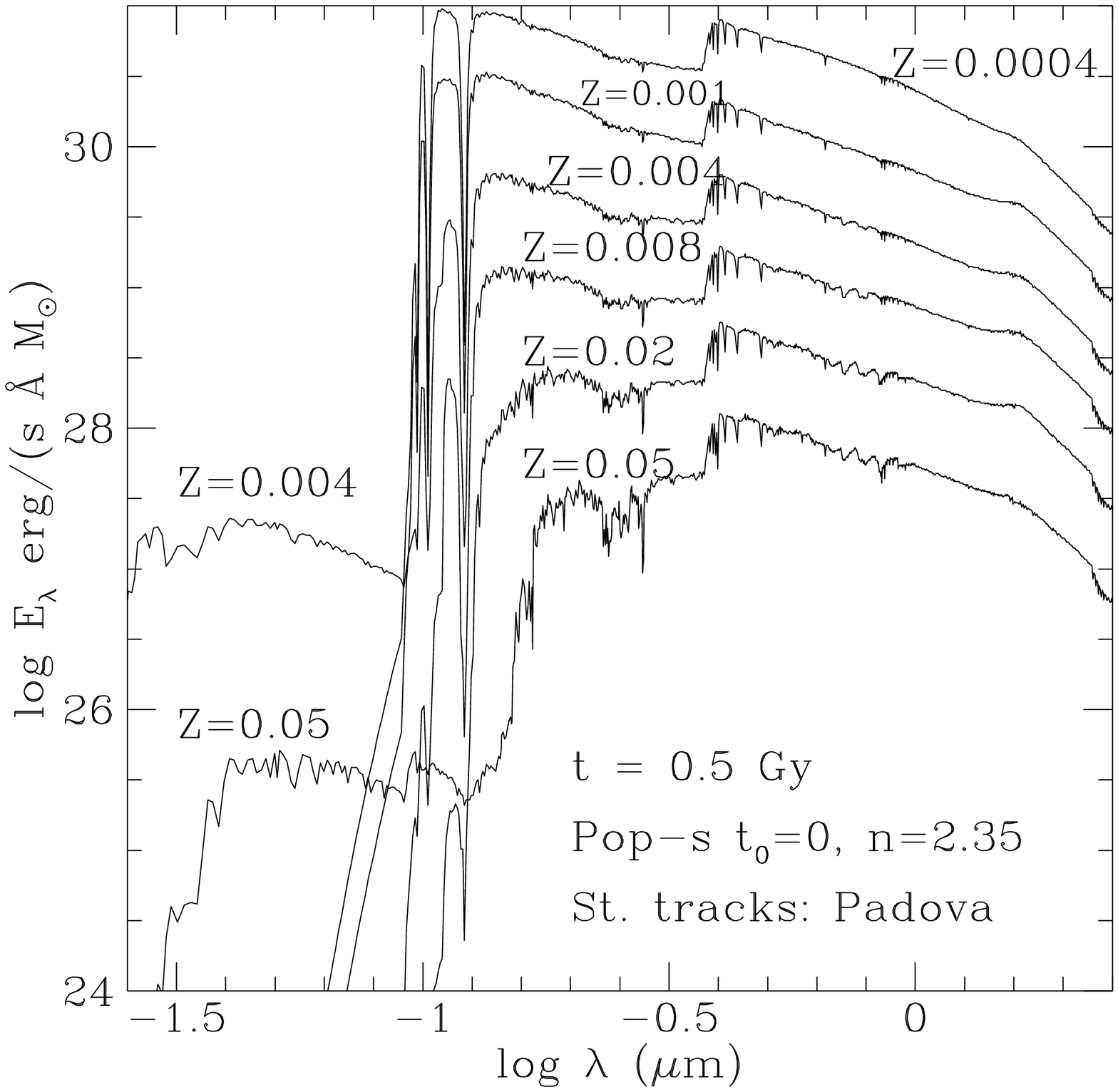}{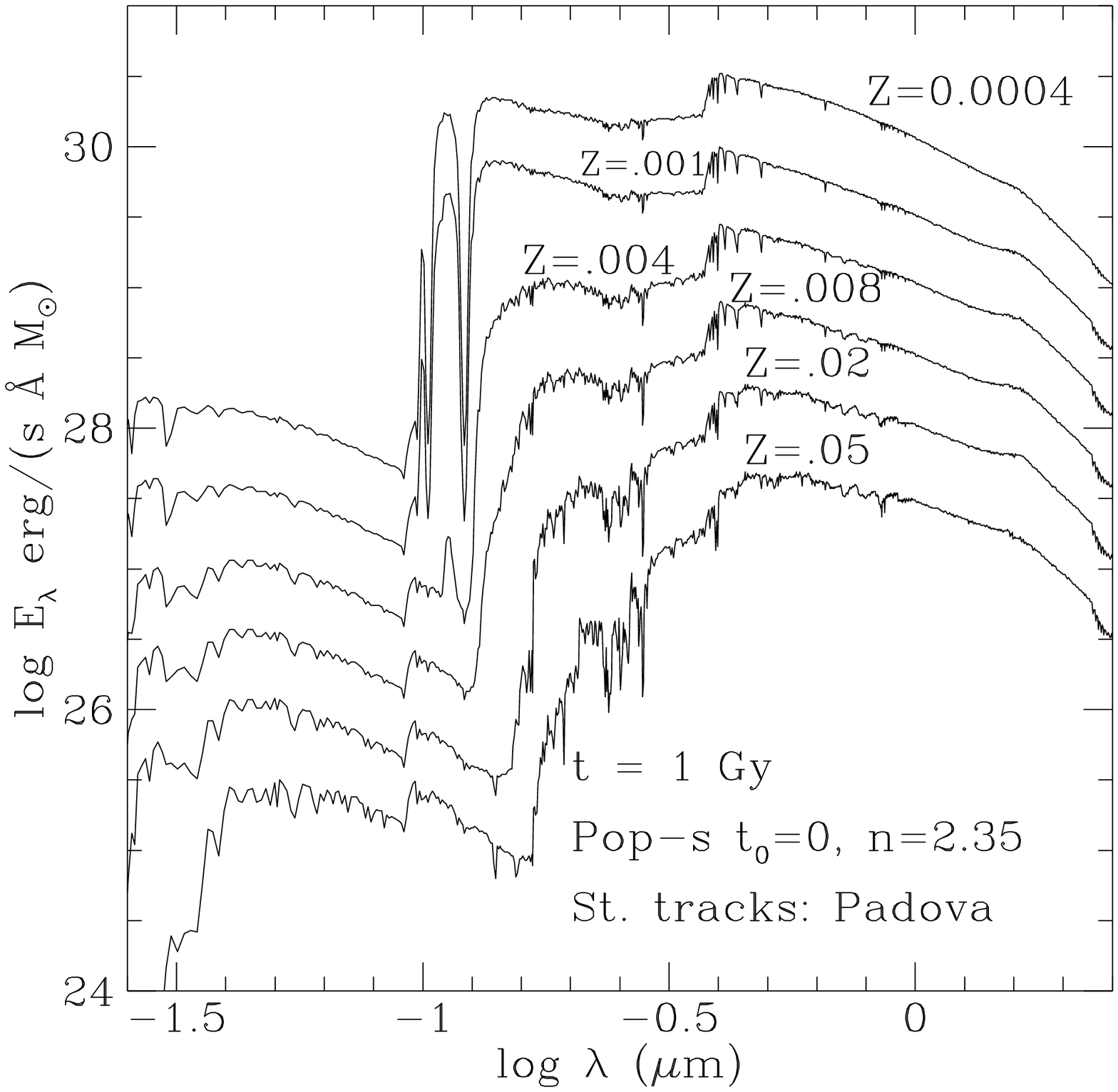}
\caption{Evolution of spectra of initial burst stellar
populations with   different metallicities. The $Z=0.0004$ spectrum has a
correct location in  log $E_\lambda$, all the others have shifted downwards by
additional 0.5  dex in respect of the previous one, with maximum shift $-2.5$
in the $Z=0.05$ case.}
\label{fig-1}
\end{figure*}

\begin{figure*}
\plottwo{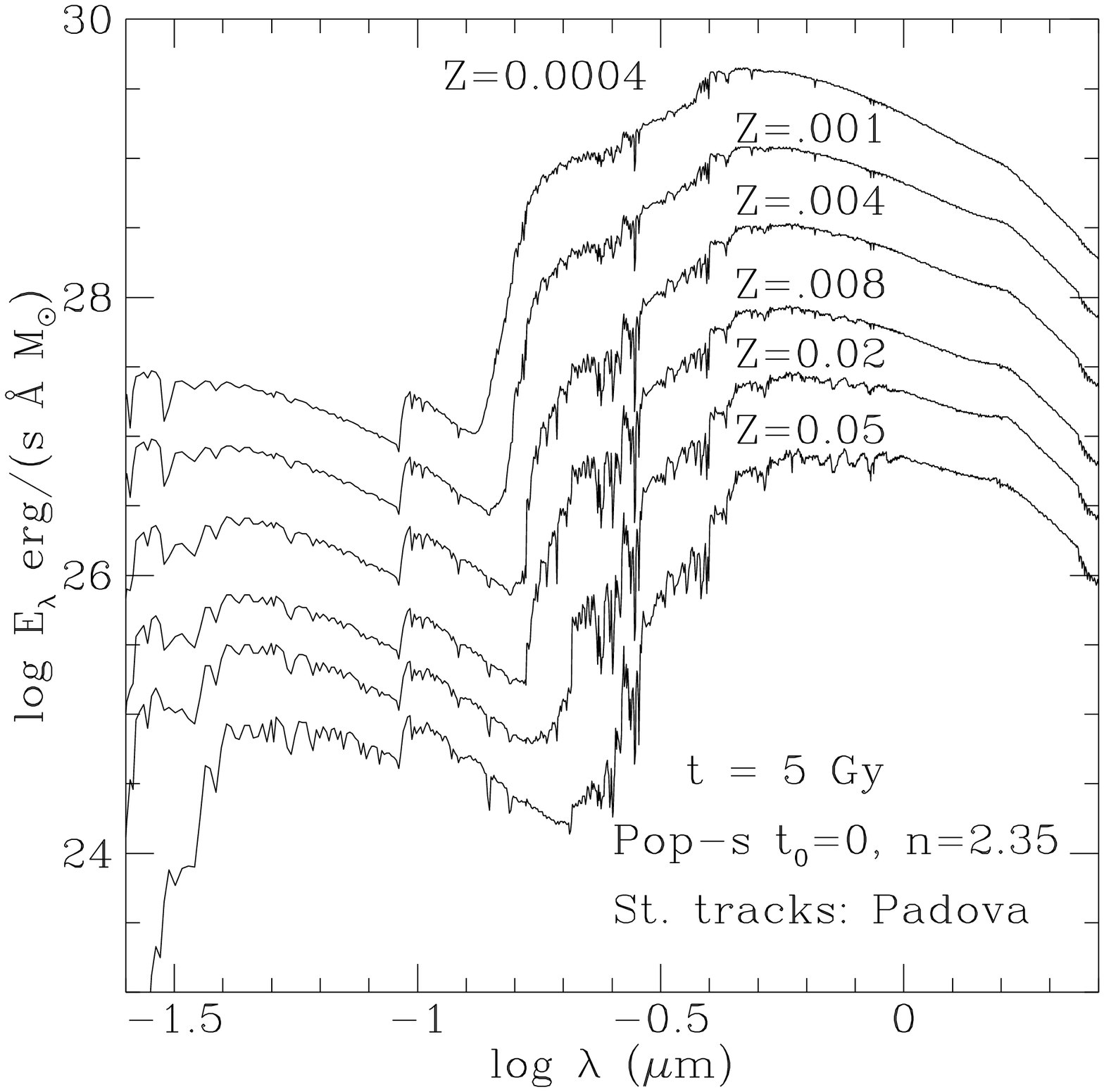}{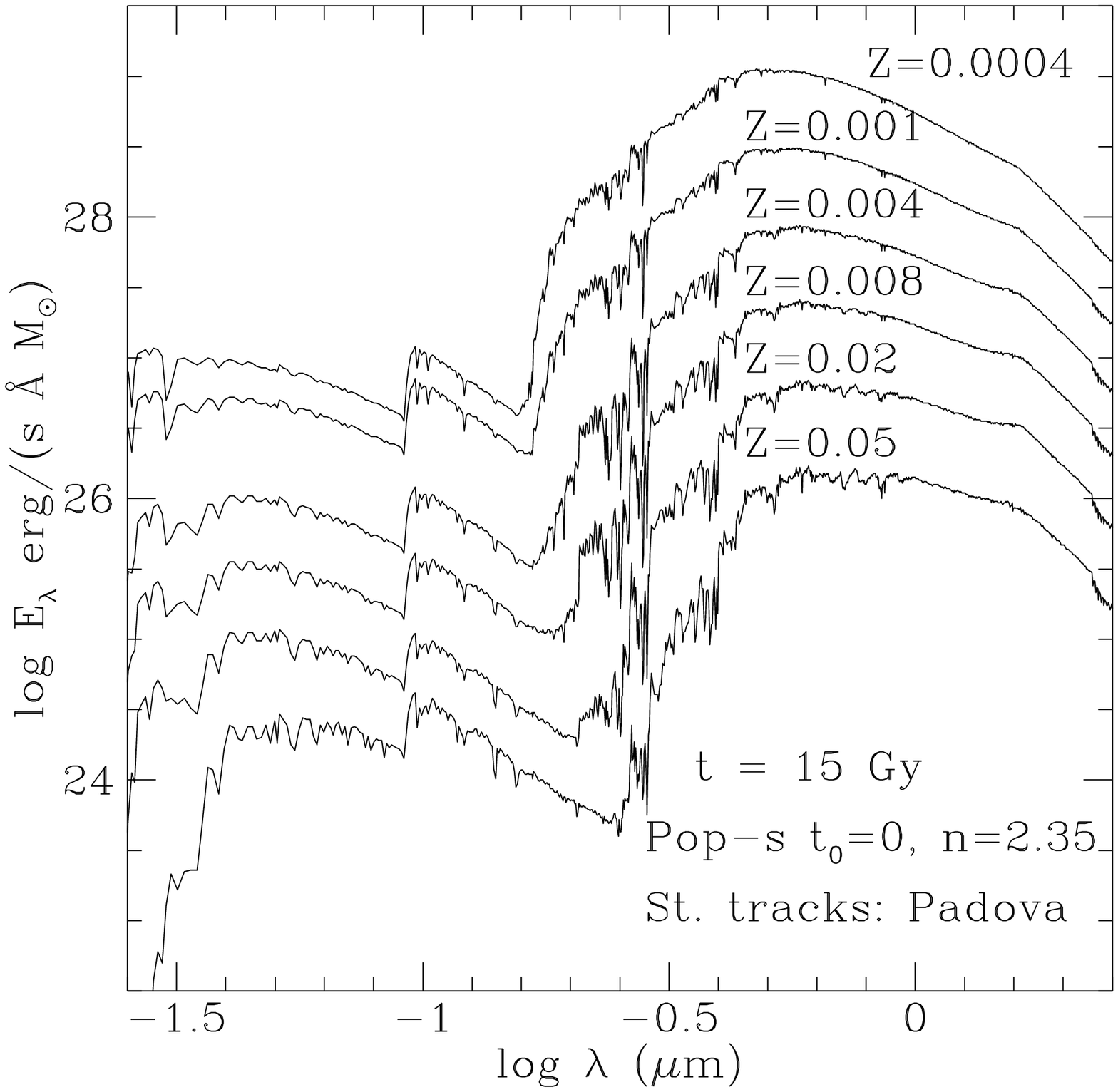}
\caption{{\it (continuation of Fig.~1)}. Spectra of the same initial
burst stellar populations at very old ages 5 and 15 Gy. Shifts on graphs 
are as indicated  in caption to Fig.~1.}
\label{fig-2}
\end{figure*}

\section{Discussion of the metallicity-dependent spectral data}
\nobreak

In the context of model stellar populations and their comparison with
observations, they fall broadly into two cathegories, the passive (initial
starburst) and active starforming populations,  the first formed in
simultaneous burst and lacking  any subsequent star
formation, the second characterized by ongoing star formation.
Later evolution of passive populations is just pure  stellar physics except
the  number distribution of stars by mass (IMF), so under
assumption of perfectness of our knowledge of stellar evolution 
it depends solely on chemical composition and IMF parameters. Active 
populations, being a superposition of infinite number of passive ones, 
additionally need for their characterization the SFR with its parameters. 

Spectral evolution of passive, initial-burst populations computed with six
compositions available in the Padova stellar tracks grid is  represented on
Fig-s~\ref{fig-1}-\ref{fig-2}. The method used for their derivation  consists
of the integration of their output over stellar mass along the  continuous
isochrones, also, to ensure a perfect "coherence" in metallicity  Kurucz model
atmosphere flux tables were interpolated to the values of  actual track set
metallicities before assigning spectra to the stars. 

\begin{figure*}
\plottwo{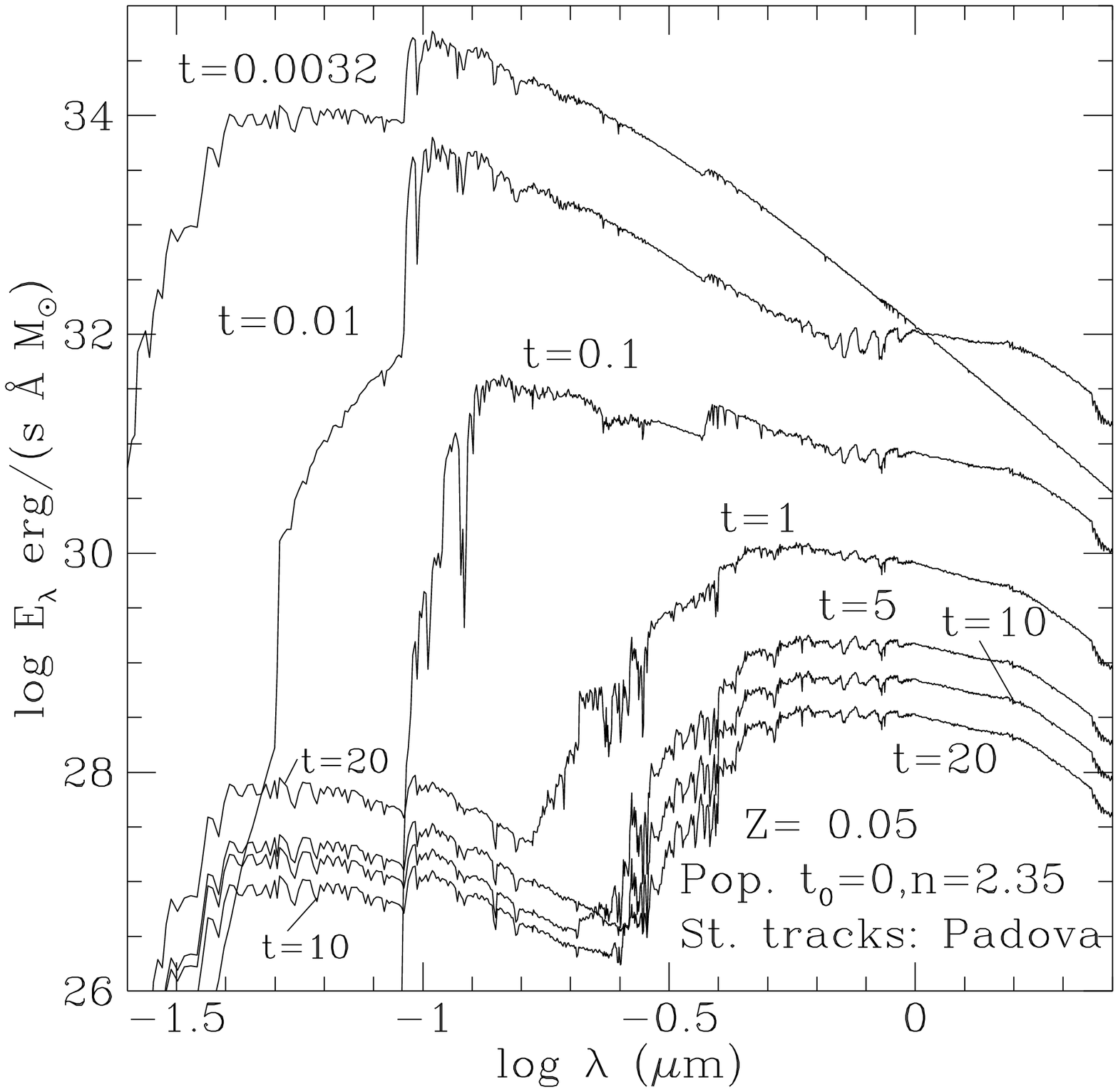}{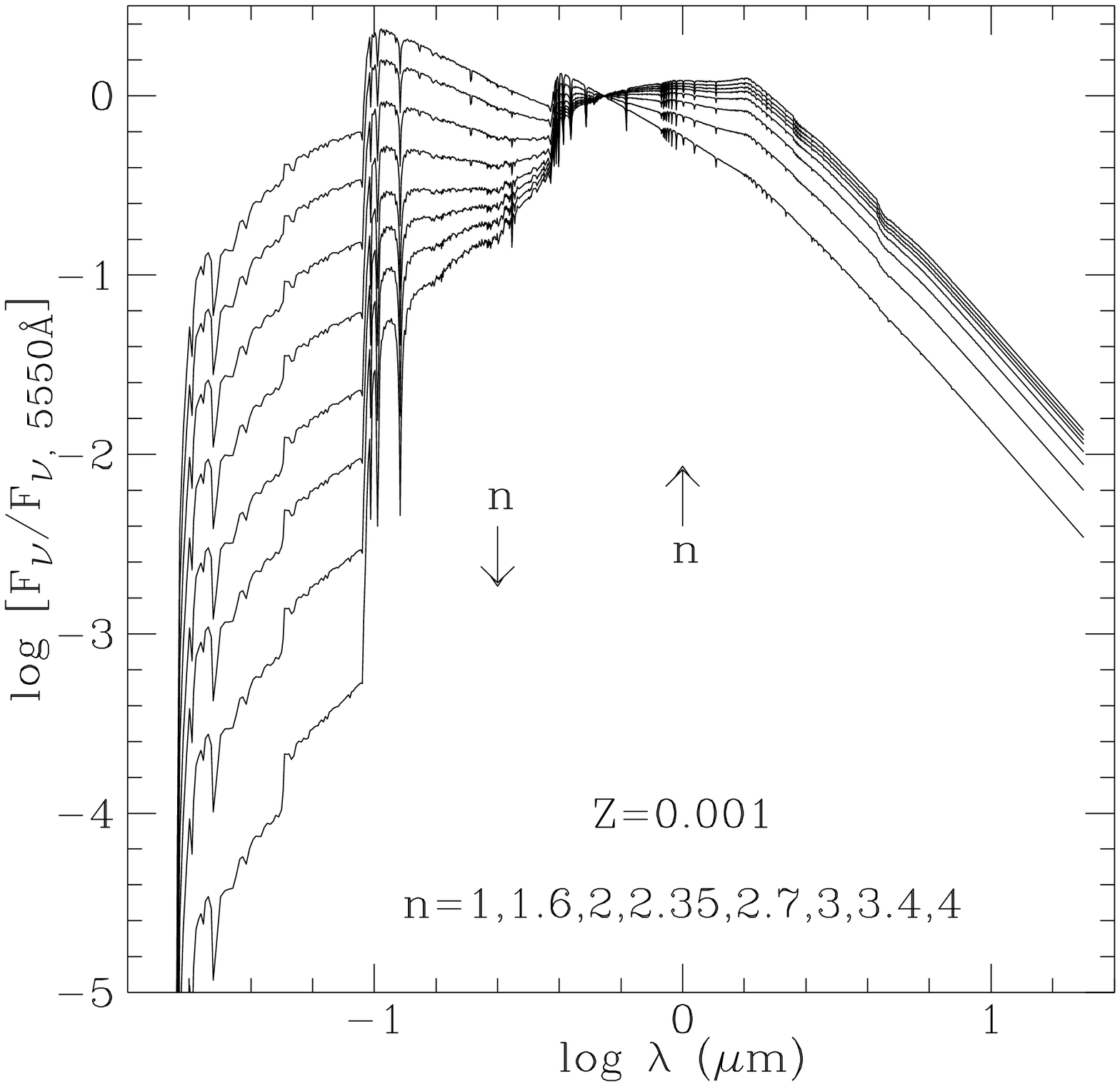}
\caption{{\it Left panel:} Age dependency of high-metallicity
initial-burst population  spectrum. \ {\it Right panel:} Impact of the IMF
slope $n$ on the total  radiation energy output of a stellar population,
integrated over its   lifetime. Given panel is based on the Geneva tracks,
distributions are   normalized at 5550 \AA. Arrows indicate the direction of
growth of $n$,   as to facilitate identification of curves.}
\label{fig-3}
\end{figure*}

General pattern of the behaviour of the passive populations is as follows. 
Were we having included a pre-main-sequence stage of stellar evolution 
with time-count set-in at the Hayashi border, with the result of massive 
stars reaching the main sequence shortly after formation and less massive 
ones only in 0.1$\div$1 Gy, we would have started the population with 
lower initial blue luminosities, with their maximal value reached the later 
the greater the IMF slope, $n$. As it is not the case here, the populations 
considered have a maximum luminosity at all wavelengths at the onset, its 
absolute value dropping with the lowering of the number of massive stars in 
the IMF($\equiv$ increase of $n$) and the monotonous decline of luminosity 
with time. The influence of $n$ on the spectrum is the most easily
understandable factor, since its growth causes the mechanical decline of 
the massive end of the IMF and, respectively, the overall mechanical 
decrease of the spectrum level towards shorter wavelengths and rise 
towards infrared, with the midst of the optical region being the approximate 
borderline (Fig.~\ref{fig-3}b). Both $Z$ (Fig-s~\ref{fig-1}-\ref{fig-2}) and
the age  (Fig.~\ref{fig-3}a) work in the same direction, the short wavelength region 
becoming progressively eroded with their growth, and the more, the shorter 
the wavelength. However, with the rise of $Z$ and the enhancement of
absorption in metallic lines in UV the bulk of energy reradiated in the 
optical and near-infrared is increasing with the definite net result that 
the spectrum level at $\lambda \geq 1$ $\mu$m is for all ages progressively
higher for higher metallicities. So a sufficiently long baseline for 
spectral or multicolor photometry coverage (plus some improvement of 
precision of both theory and observations) could probably resolve the 
age-metallicity degeneracy, I think a theoretical calibration might be 
attempted after the release of improved cool-star atmospheres. 

There are yet some more noteworthy details pointed out on
Fig-s~\ref{fig-1}-\ref{fig-2} like the extremely rapid time evolution of the
He\raisebox{.6ex}{$\circ$}  continuum below 504\AA   and bit shallower change
of the Lyman jump, making  both of them theoretically good age estimators for
the extremely young  stellar populations. Fig.~\ref{fig-3}a is also embedding
a reference to a  specific peculiarity of the Padova track system: the
high-metallicity  population includes a hot AGB-{\it manque} ($\equiv$naked
stellar core)  stellar evolution phase, stars in this stage are the cause of
the UV-flux  upturn at ages $\geq$ 10 Gy. This interesting result is in full
detail  covered by the authors of tracks (Bressan et al. 1994).

\begin{figure*}
\plottwo{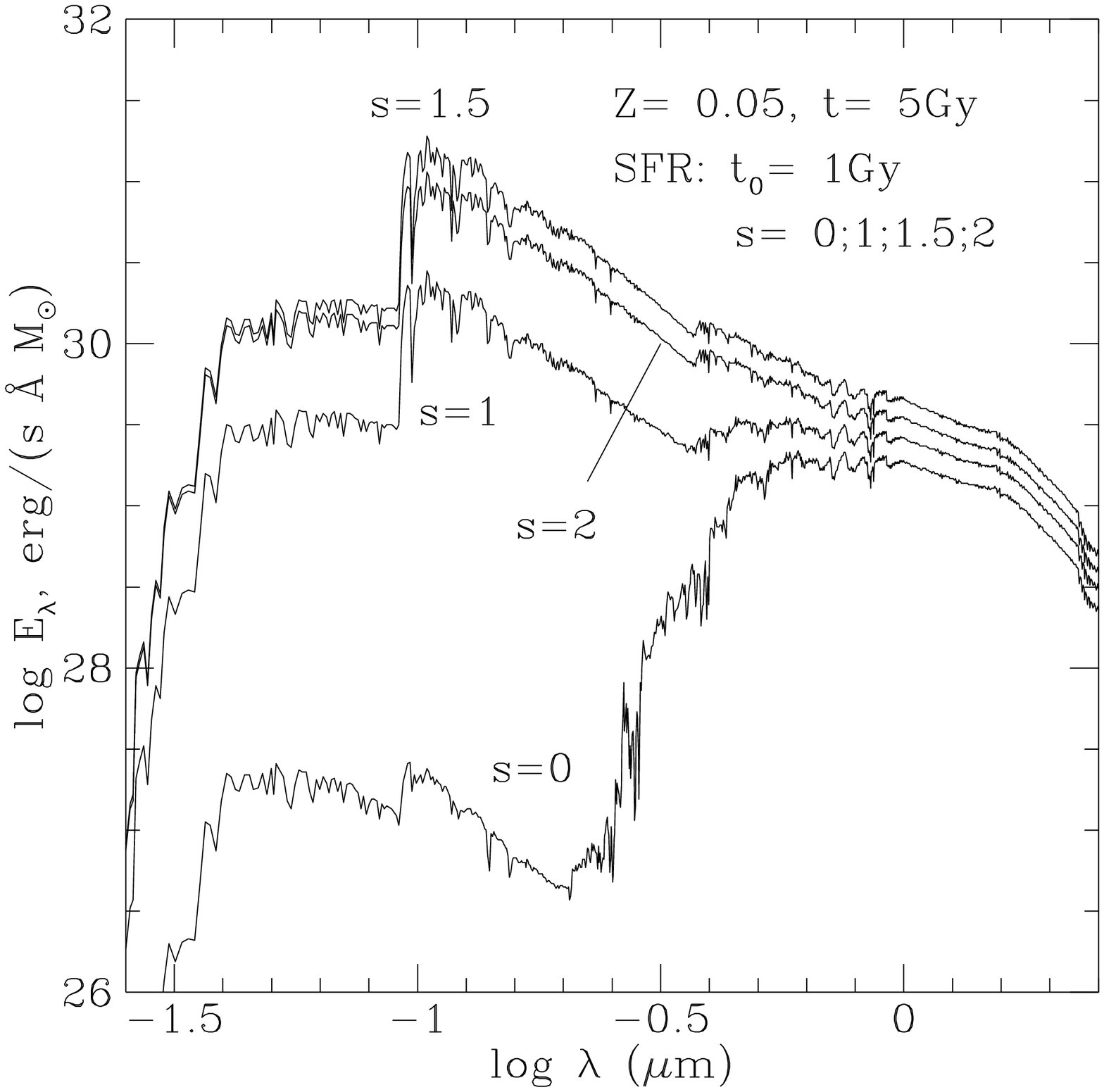}{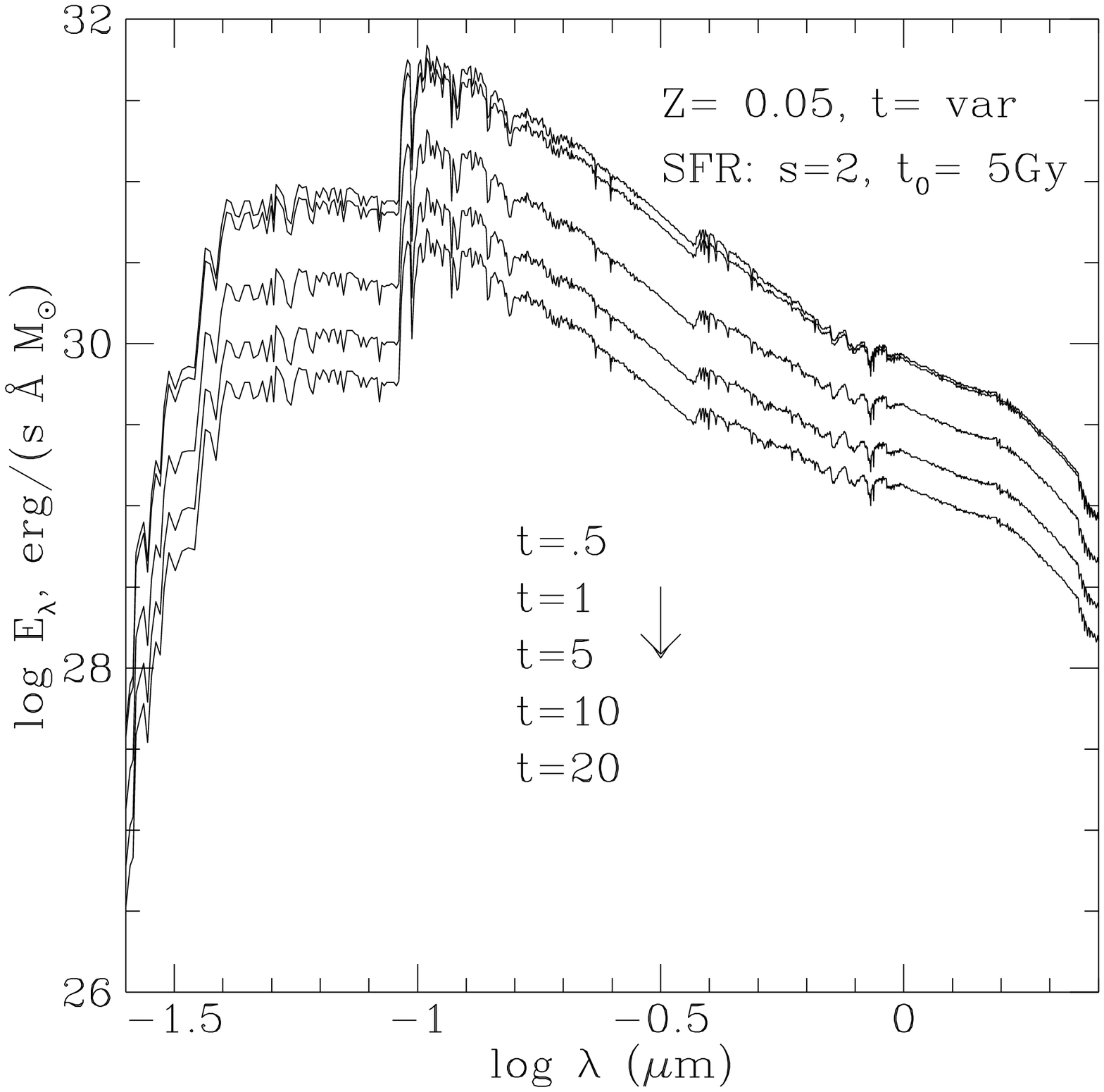}
\caption{{\it Left panel:} Spectral evolution of a stellar population 
with continuing star-formation, having time-scale $t_0=5$. In the given 
s=2 case half of the matter will during $t_0$ be converted to stars. \ 
{\it Right panel:} Influence of the SFR time behaviour differencies, as 
introduced by its power index $s$ values, on the spectrum of a $Z=0.05$ 
 population with continuous star formation.}
\label{fig-4}
\end{figure*}

Spectral evolution of active populations is governed to a large extent by 
the SFR shape (i.e. power index $s$ in the case here) and star-formation 
timescale, $t_0$. The reddest limiting case for both the spectrum and 
colors being the passive $t_0$=0 population, the active population with the 
given $s$ forms during its youth stars the slower the larger $t_0$ value is, 
but has higher star-formation intensity and bluer colors at later epochs. 
The limits to what extent the selection of the functional form parameter 
$s$ may influence the spectrum are approximately the same, the dependence 
of spectral evolution on $s$ being for one parameter set illustrated on 
Fig.~\ref{fig-4}a ($s$=0 $\equiv$constant, $s$=1 $\equiv$ exponentially 
declining star formation rate). The scale of impact depends also on the 
age of the population, in the $s$=0, $t_0$=1 Gy model the star formation 
has had an duration of first 1 Gy and not been working for 4 Gy-s until 
our time of observation, it makes the population spectra red and close 
to that of a passive population at the same age, 5 Gy. Right panel depicts 
the temporal evolution of an active stellar population with a quite extended 
timescale, so although both its colors and spectrum are slowly evolving 
to red, no impressive changes next to its overall absolute decrease are 
evident, even the UV-branch keeping nearftp ly the same level in respect to 
the optical region, as a result of the continuing star formation at the 
levels not drastically differing from those in earlier stages.  

\section{Metallicity and radiation output in specific wavelength ranges}

\begin{table*}
\vspace*{-.4cm}
{\bf Table 1. Time-integrated energy fluxes} of coeval stellar populations
with different IMF slopes and metallicities, per 1 M$_\odot$ in massive 
stars $10 \div 120$  M$_\odot$.  
Flux units  $10^{52}$ erg, flux fractions f$_{\rm i}$ computed for 
intervals 90.9-395 \AA, 395-995 \AA, 995-1995 \AA, 1995-3590 \AA, 
3590-7490 \AA, 0.749-4.99 $\mu$m and 4.99-160 $\mu$m. \hfill
\vspace{.8ex}
\centering{\small
\begin{tabular}{c|c|cccccccc} 
 \hline 
\mbox{\rule[-.9ex]{0ex}{3.2ex} n} & Z & L & f$_1$ & f$_2$ & f$_3$ & f$_4$ & f$_5$ & f$_6$ & f$_7$ \\ 
\hline 
\mbox{\rule{0ex}{2.2ex} 1.0}  & 0.001 &  0.2944 &.0792 &.2585 &.4691 &.1090 &.0635 &.0206 &.0002\\
     & 0.004 &  0.2977 &.0503 &.2429 &.4932 &.1153 &.0695 &.0284 &.0002\\
     & 0.008 &  0.2896 &.0383 &.2465 &.5128 &.1170 &.0590 &.0262 &.0003\\
     & 0.020 &  0.2629 &.0243 &.2408 &.5265 &.1172 &.0539 &.0368 &.0005\\
     & 0.040 &  0.2305 &.0096 &.2012 &.5476 &.1367 &.0659 &.0385 &.0005\\
\hline 
1.6  & 0.001 &  0.3169 & .0579 & .2025 & .4809 & .1239 & .0911 & .0434 & .0004\\
     & 0.004 &  0.3147 & .0365 & .1902 & .4896 & .1263 & .0988 & .0579 & .0006\\
     & 0.008 &  0.3074 & .0271 & .1880 & .5032 & .1287 & .0919 & .0603 & .0007\\
     & 0.020 &  0.2824 & .0166 & .1799 & .5020 & .1256 & .0888 & .0859 & .0012\\
     & 0.040 &  0.2487 & .0067 & .1527 & .5175 & .1412 & .0962 & .0844 & .0012\\
\hline 
2.0  & 0.001 &  0.3897 & .0376 & .1405 & .4561 & .1388 & .1393 & .0868 & .0009\\
     & 0.004 &  0.3746 & .0239 & .1339 & .4546 & .1365 & .1444 & .1055 & .0012\\
     & 0.008 &  0.3678 & .0174 & .1296 & .4578 & .1375 & .1420 & .1144 & .0014\\
     & 0.020 &  0.3458 & .0103 & .1201 & .4352 & .1291 & .1419 & .1611 & .0023\\
     & 0.040 &  0.2997 & .0043 & .1052 & .4519 & .1415 & .1441 & .1509 & .0021\\
\hline 
2.2  & 0.001 &  0.4691 & .0274 & .1064 & .4253 & .1459 & .1741 & .1197 & .0013\\
     & 0.004 &  0.4399 & .0176 & .1027 & .4200 & .1411 & .1772 & .1397 & .0016\\
     & 0.008 &  0.4337 & .0127 & .0982 & .4162 & .1403 & .1775 & .1533 & .0019\\
     & 0.020 &  0.4159 & .0073 & .0888 & .3811 & .1278 & .1778 & .2141 & .0030\\
     & 0.040 &  0.3538 & .0032 & .0799 & .4000 & .1388 & .1782 & .1973 & .0027\\
\hline 
2.35 & 0.001 &  0.5620 & .0205 & .0823 & .3947 & .1502 & .2029 & .1477 & .0016\\
     & 0.004 &  0.5159 & .0133 & .0805 & .3872 & .1436 & .2049 & .1684 & .0019\\
     & 0.008 &  0.5106 & .0095 & .0762 & .3779 & .1411 & .2070 & .1861 & .0023\\
     & 0.020 &  0.4985 & .0053 & .0674 & .3346 & .1249 & .2063 & .2578 & .0037\\
     & 0.040 &  0.4166 & .0024 & .0621 & .3547 & .1350 & .2066 & .2359 & .0032\\
\hline 
2.5  & 0.001 &  0.6974 & .0148 & .0613 & .3599 & .1532 & .2319 & .1769 & .0020\\
     & 0.004 &  0.6256 & .0097 & .0607 & .3505 & .1452 & .2336 & .1980 & .0023\\
     & 0.008 &  0.6221 & .0068 & .0568 & .3360 & .1405 & .2372 & .2201 & .0027\\
     & 0.020 &  0.6198 & .0037 & .0490 & .2864 & .1204 & .2340 & .3021 & .0043\\
     & 0.040 &  0.5075 & .0017 & .0464 & .3071 & .1298 & .2355 & .2757 & .0037\\
\hline 
2.7  & 0.001 &  0.9767 & .0090 & .0391 & .3115 & .1551 & .2681 & .2148 & .0024\\
     & 0.004 &  0.8492 & .0060 & .0394 & .2997 & .1458 & .2706 & .2358 & .0027\\
     & 0.008 &  0.8506 & .0041 & .0363 & .2793 & .1377 & .2754 & .2640 & .0032\\
     & 0.020 &  0.8719 & .0022 & .0303 & .2255 & .1126 & .2666 & .3577 & .0050\\
     & 0.040 &  0.6937 & .0010 & .0297 & .2451 & .1211 & .2717 & .3270 & .0044\\
\hline 
3.0  & 0.001 &  1.7551 & .0038 & .0183 & .2447 & .1540 & .3122 & .2641 & .0030\\
     & 0.004 &  1.4602 & .0026 & .0188 & .2295 & .1437 & .3183 & .2839 & .0032\\
     & 0.008 &  1.4817 & .0017 & .0169 & .2035 & .1306 & .3229 & .3206 & .0038\\
     & 0.020 &  1.5820 & .0009 & .0133 & .1510 & .0993 & .3025 & .4269 & .0060\\
     & 0.040 &  1.2094 & .0004 & .0138 & .1660 & .1063 & .3152 & .3931 & .0052\\
\hline 
3.4  & 0.001 &  4.2012 & .0011 & .0060 & .1768 & .1485 & .3510 & .3130 & .0036\\
     & 0.004 &  3.3257 & .0007 & .0063 & .1585 & .1382 & .3636 & .3290 & .0037\\
     & 0.008 &  3.4400 & .0005 & .0054 & .1303 & .1191 & .3651 & .3752 & .0044\\
     & 0.020 &  3.8395 & .0002 & .0040 & .0867 & .0828 & .3286 & .4907 & .0069\\
     & 0.040 &  2.8204 & .0001 & .0044 & .0948 & .0875 & .3505 & .4567 & .0060\\
\hline 
4.0  & 0.001 & 16.8714 & .0002 & .0010 & .1147 & .1383 & .3811 & .3604 & .0043\\
     & 0.004 & 12.6020 & .0001 & .0011 & .0950 & .1293 & .4026 & .3678 & .0041\\
     & 0.008 & 13.4144 & .0001 & .0009 & .0693 & .1037 & .3968 & .4243 & .0050\\
     & 0.020 & 15.6599 & .0000 & .0006 & .0396 & .0647 & .3416 & .5457 & .0078\\
     & 0.040 & 11.1299 & .0000 & .0007 & .0415 & .0662 & .3711 & .5137 & .0067\\
\hline
\end{tabular}}
\end{table*}

Metallicity growth affects the stellar evolution in two ways: first, making 
stars cooler and dimmer, secondly, redistributing their flux through opacity growth 
at short wavelengths towards longer wavelengths. 
So the summatic result of $Z$ rise on the composite spectrum of a stellar population 
is the progressive erosion of flux in ultraviolet region, and the faster, the 
shorter the wavelength. However, with the growth of $Z$ and the
enhancement of absorption in metallic lines in UV the bulk of energy
reradiated in the  optical and near-infrared is increasing with the definite
net result that  the spectrum level at $\lambda \geq 1$ $\mu$m is for all
population ages  progressively higher for higher metallicities. 

Table 1 gives a general quantitative review of the extent of metallicity 
effects in different spectral regions for populations with different mass
function slopes $n$. Time is eliminated by integration over the lifetimes of 
stars,  population mass is scaled to the unit amount of mass in very massive 
stars $M\ge 10$ \msun, Geneva set of tracks (5 compositions, the 
$Z=0.10$ subset was not included) was used. These data testify, that 
composition-caused flux changes can be rather impressive, extending to factor 
of 10 in far-UV and $\sim 2$ in infrared. 

\begin{figure*}
\plottwo{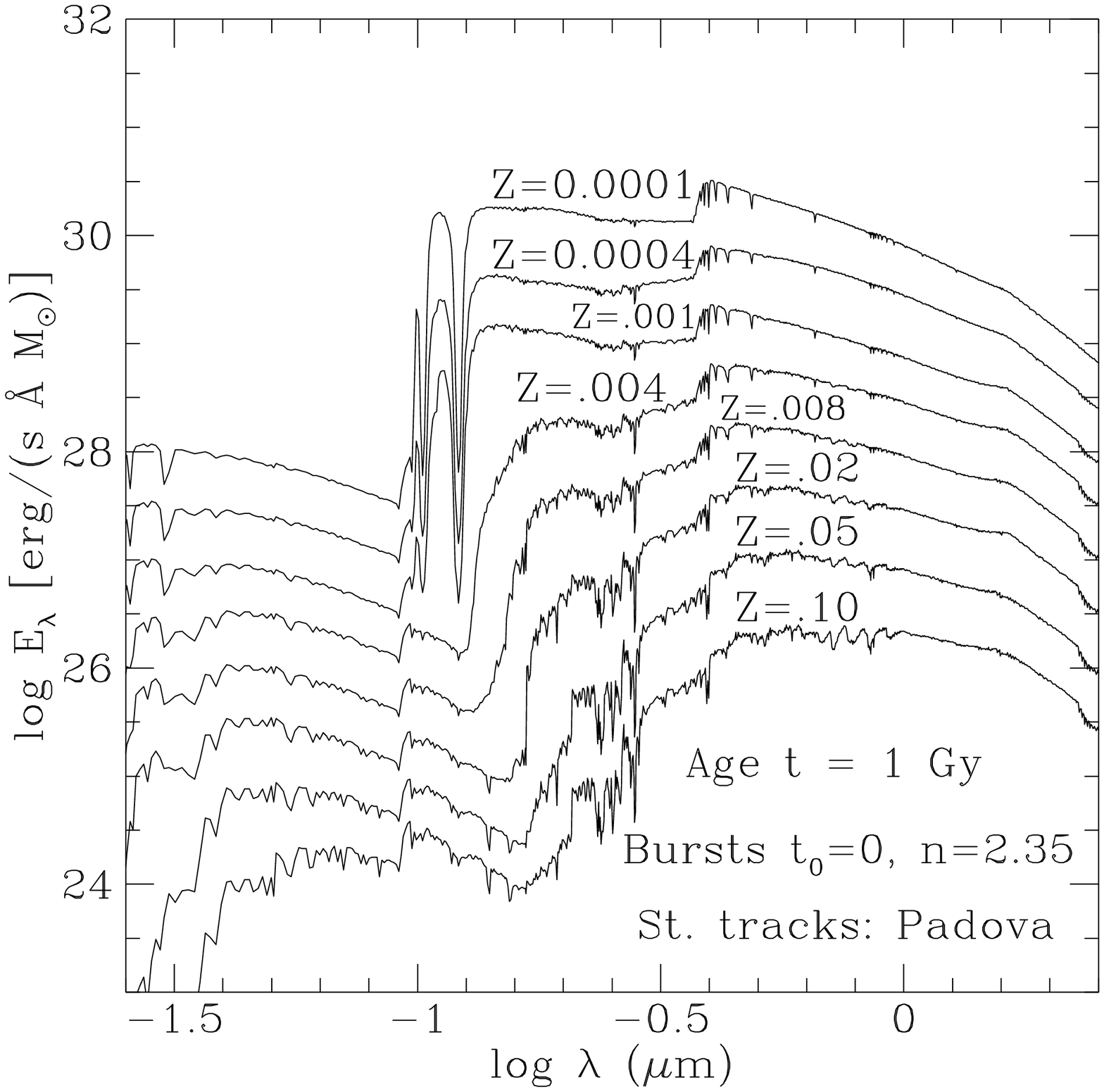}{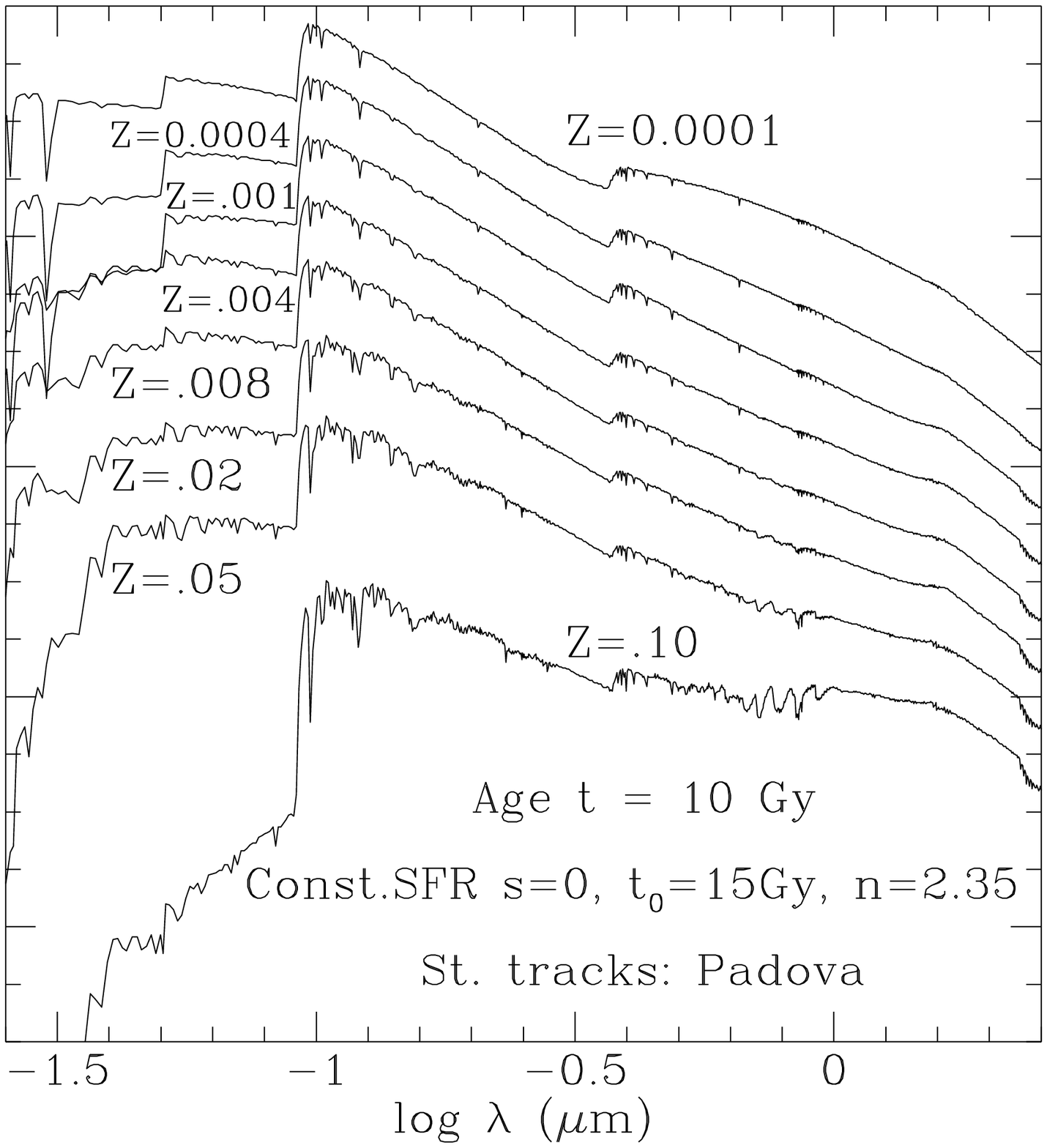}
\caption{Composition dependence of spectra of stellar populations: coevally formed 
young stellar populations (age 1 Gy, {\it left panel}) and old populations with constant SFR
(age 10 Gy, {\it right panel}).}
\label{fig-5}
\end{figure*}

As to illustrate metallicity effects graphically, we also provide a couple of 
examples on Fig.~\ref{fig-5}, computed with Padova tracks. On the left panel
of this figure  we have plotted computed spectra 
of coevally formed stellar populations (so-called {\it initial burst}  populations) 
with 8 different initial metallicities $Z$, having the same ages 1 Gy. Such
comparatively blue spectra  are typical for medium-aged 
star clusters of assumed chemical compositions,    
or might be relevant to different parts inside young elliptical galaxies with
ages somewhat exceeding 1 Gy.  The flux of models is scaled 
to the unit mass in luminous stars with masses $M > 0.6$ \msun, 
the IMF in these graphs is a power-law with slope $n = -2.35$ 
("Salpeter" value). The most metal-deficient, $Z=0.0001$ spectrum has 
a correct location in log $E_\lambda$, all the others have been 
successively shifted downwards by 
additional 0.5 dex, with maximum total shift $-3.5$ 
for the $Z=0.10$ case. 
On the right panel
the spectra of old, 10 Gy stellar populations, are plotted, in which star formation 
is continuous and proceeds
 with a constant, time-independent intensity over the 
eon $t_0$. Given case might be considered as a kind of approximation to late spirals 
or irregulars,
since in many of these actual SFR-s do not seem to 
significantly differ from the mean average over their past. 
Notice, however, that due to 
the actively continuing star formation the ultraviolet flux of models keeps 
 sizable. The growth of metal content sharply reduces the flux at shorter 
wavelengths, as also on the left-panel plot. 

\section{Nucleosynthesis}
The tabulated nucleosynthesis of He, C and O will be available in replacement
paper to this first draft!

\end{document}